# Scintillation Characterizations of $Tl_2LiLuCl_6$: $Ce^{3+}$ Single Crystal

Gul Rooh[a], H. J. Kim[b*], Jonghoon Jang[b], and Sunghwan Kim[c]

[a]Department of Physics, Abdul Wali Khan University, Mardan, 23200, Pakistan
[b]Department of Physics, Kyungpook National University, Daegu 702-701, Korea
[c]Department of Radiological Science, Cheongju University, Cheongju 360-764, Korea

[*] Corresponding author.
Tel.:+82-53-950-5323; fax: +82-53-956-1739. E-mail address: hongjoo@knu.ac.kr (H.J. Kim).

**Abstract**

0.5%, 1%, 3% and 5% Ce-concentration single crystals of $Tl_2LiLuCl_6$ were grown from the melt using two zone vertical Bridgman technique. X-ray induced emission spectra showed $Ce^{3+}$ emission between 370 nm and 540 nm wavelength range. Energy resolution, light yield and decay time of the grown samples were measured under γ-ray excitation at room temperature. Energy resolution of 5.6% (FWHM) with 27,000±2700 light yield is found for 1%Ce doped sample. For the same dopant concentration, three decay time components are also observed. Variation of scintillation properties is observed as a function of dopant concentration in this material. This material will provide excellent detection efficiency for X- and γ-rays due to its high effective Z-number and density. It is expected that this scintillor will be a potential detector for the medical imaging techniques.

**Introduction**

Thallium (Tl) doped NaI and CsI scintillators [1, 2] are widely used in different applications for the detection of X- and γ-rays. Tl possesses various scintillation properties which make this ion very interesting for the development of single crystal. For example, in CsI and NaI scintillators, Tl ion acts as luminescence center; while due to its high Z-number and density it is a best choice for the applications where high density and Z-number compounds are required. Recently, we have discovered and reported new Tl-based alkali halide elpasolites [3-5]. Among these scintillators, $Tl_2LiGdCl_6$: $Ce^{3+}$ [3] and $Tl_2LiYCl_6$: $Ce^{3+}$ [4] shows excellent scintillation performance which includes good energy resolutions of 4.8% (FWHM), and 4.6% (FWHM) with 58,000 ph/MeV and ~31,000 ph/MeV light yields, respectively at room temperature under the γ-ray excitation. In addition, the effective Z-numbers of these scintillators are higher than that of LYSO: $Ce^{3+}$, NaI: Tl, CsI: Tl and $LaBr_3$: $Ce^{3+}$. Also, considering the presence of lithium in the host lattice, density and high Z-number of $Tl_2LiGdCl_6$: $Ce^{3+}$ and $Tl_2LiYCl_6$: $Ce^{3+}$, it is expected that these scintillators could be used in the fields of medical imaging, homeland security and neutron detection.

In the present study we investigated the scintillation properties of our newly discovered Ce-activated $Tl_2LiLuCl_6$ (TLLC) scintillation material. Scintillation properties were studied under X- and γ-ray excitation at room temperature. Comparing with $Cs_2LiLuCl_6$, the effective Z-number and density of $Tl_2LiLuCl_6$ is higher due to the greater Z-number and density of Tl than that of Cs.

**Experimental Technique**

*A. Crystal growth*

$Tl_2LiLuCl_6$: $Ce^{3+}$ single crystals of various Ce-concentrations were grown by two zone vertical Bridgman technique. Starting materials (TlCl, LiCl, $LuCl_3$ and $CeCl_3$ of ~5N purity powders) from Sigma-Aldrich or Alfa-Aesar were weighing in ultra-dry argon purged glove box having moisture and oxygen level less than 5 ppm. Weighing powder was loaded in ultra-clean quartz ampoule having an inner diameter of 8 mm. Loaded quartz ampoule with the powder were sealed with an Oxy-propane torch under dynamic vacuum of ~$10^{-7}$ Torr. After sealing, all the ampoules were lowered into two zone vertical Bridgman chamber with the help of a synchronous motor. Temperature of the furnace was raised to 570 °C, i.e. 40 °C higher than the melting point of TLLC. A temperature gradient of 10°C/cm was achieved by adjusting the upper and lower zone temperatures of the Bridgman furnace. During the crystal growth process, the rate of growth was maintained at 10-12 mm/day. Finally, the grown samples were transparent and small size samples with the dimensions of ~$\phi$8 x 1mm$^3$ were cut and polished for the scintillation characterization. Figure 1 show pictures of the as grown samples.

### B. Equipment

Emission spectra of TLLC crystals under X-ray excitation were obtained at room temperature. Investigated sample were irradiated with an X-ray tube (DRGEM. Co.) using a tungsten anode operating at 50 kV and 1 mA. The emission spectra were measured by utilizing a spectrometer (QE65000 fiber optic spectrometer) made by Ocean Optics. Pulse height spectra were measured with a Hamamatsu R6233 photomultiplier tube (PMT) at room temperature. Polished sample crystals were wrapped in several layers of 0.1-mm-thick UV reflecting Teflon tape except one face which was coupled directly to the entrance window of the PMT using index matching optical grease. After irradiation with γ-rays from a $^{137}$Cs source, the analog signals produced in the crystal were shaped with a Tennelec TC 245 spectroscopy amplifier and were fed into a 25-MHz flash analog-to-digital converter (FADC). A software threshold was set to trigger an event by using a self-trigger algorithm on the field programmable gate array (FPGA) chip on the FADC board. The FADC output was recorded into a personal computer by using a USB2 connection, and the recorded data are analyzed with a C++ data analysis program [6].

Decay time spectra was measured after optically coupled crystals of TLLC: $Ce^{3+}$ with the PMT (Hamamatsu R6233) and was irradiated by 662 keV γ-rays from a $^{137}$Cs source. PMT signals were fed into a 400-MHz FADC and a home-made FADC module is fabricated to sample the pulse every 2.5 ns for duration up to 64 µs so that one can fully reconstruct each photoelectron pulse [7]. From the recorded pulse shape information, the decay time of TLLC: $Ce^{3+}$ was obtained.

## Scintillation properties

### A. Crystal analysis

TLLC was reported by Mayer et al. [8] and has a tetragonal crystal structure. It has P4/*nbm* space group with lattice constants a = 10.145 Å and c = 10.251 Å. The volume and density of the unit cell are obtained as 1055.0 Å$^3$ and 5.06 g/cm$^3$, respectively. Effective Z-number of this material is found to be 71.

### B. X-ray excited luminescence

X-ray induced luminescence spectra of the Ce-activated TLLC single crystals are displayed in Fig. 2. The observed emission is attributed to the parity allowed *5d → 4f* transition of the $Ce^{3+}$ ion, i.e. transition from the lowest *5d* level to $^2F_{5/2}$ and $^2F_{7/2}$ levels of the $4f^1$ configuration [9]. The observed emission spectra are not identical and showed a shift towards higher wavelength regions with the increase of Ce-concentration in the host. For 0.5% and 1% Ce-concentration, broad emission bands are located between 370 and 540 nm, peaking at 428 nm. While 3% and 5% Ce samples shows similar emission bands between 370 and 540 nm with a slight shift of the peak maxima towards higher wavelengths i.e. 437 nm and 435 nm, respectively. Similar trends in the wavelength shift is also reported in different alkali halide compounds [10, 11]. X-ray excitation luminescence spectra of $Cs_2LiLuCl_6$: $Ce^{3+}$ showed low intensity remanent STE emission [12], similar emission was not observed in the spectra of TLLC: $Ce^{3+}$, see Fig. 2. The emission spectra of the TLLC: $Ce^{3+}$ revealed that this material is attractive for γ-ray spectroscopy, since it matches well with the response function of the photomultiplier tubes and silicon photo-diodes.

### C. Pulse height spectrum and light yield

Pulse height spectra of the Ce doped TTLC crystals are obtained under γ-ray excitation from a $^{137}Cs$ source. The energy resolutions of the samples are calculated by applying a Gaussian fit to the photopeaks after irradiation with γ-ray source. Best energy resolution of 5.6 % (FWHM) is obtained for 1% Ce doped TLLC crystal, Fig. 3 shows the pulse height spectrum of 1% Ce doped TLLC crystal at room temperature. Higher value of Ce content in the host lattice TLLC cannot improve energy resolution. Energy resolutions of the all the doped samples are presented in Table 1. In all the grown samples of TLLC a satellite peak was observed at low energy in the pulse height spectra due to Tl K-X-ray escape peak.

Light yield (LY) of the TLLC: Ce crystals are measured under γ-ray irradiation from a $^{137}Cs$ source at room temperature. During the LY measurement, similar experimental setup of pulse height measurement as mentioned in section-B is utilized. LY is evaluated by comparing the channel numbers in the pulse height spectra of the TLLC: $Ce^{3+}$ samples with that of LYSO reference crystal (absolute LY = 33,000 ph/MeV) under similar conditions of PMT bias, shaping time and amplifier gain. Maximum LY of 27,000±2700 ph/MeV is obtained for 1% Ce doped sample using a 6μs shaping time constant. Figure 4 shows the pulse height spectra of TLLC: 1% $Ce^{3+}$ and LYSO under γ-ray excitation from a $^{137}Cs$ source. Details of the obtained LY for all doped samples are presented in Table 1.

**Decay times**

Scintillation time profiles of the Ce-activated TLLC crystals are shown in Fig. 5 at room temperature. The obtained spectra are fitted with the sum of two and three exponential decay functions. Except 0.5% Ce sample, all samples show three decay time constants (shown inset Fig. 5). Decay time constants and their relative contribution to the total light yield are given in Table I. From the Table I, it is clear that the decay time constants vary with the increase of Ce-concentration in the host lattice. An increase of Ce-concentration of 0.5% to 5% made this compound slower i.e. the decay time constants get longer. Comparing the decay time constants of TLLC: $Ce^{3+}$ with TLGC: $Ce^{3+}$ [3] and TLGB: $Ce^{3+}$ [5], this scintillator shows significantly slow scintillation response at room temperature. The possible reason of slow down might be due to the inefficient energy transfer from the host lattice to the $Ce^{3+}$ ion in this material.

**Conclusion**

Scintillation properties of the newly developed Ce-activated TLLC single crystals are investigated. Comapring with the scintillators used in the Positron Emission Tomography (PET) i.e. LYSO: $Ce^{3+}$, LSO: $Ce^{3+}$ and BGO: $Ce^{3+}$ etc. , this material possessed high effective Z - number and could replace these scintillators in the future PET application. All the doped samples exhibit broad emission bands due to 5d → 4f transition of $Ce^{3+}$ ion. Among the grown samples, 1%Ce doped shows an energy resolution of 5.6% (FWHM) and 27,000±2700 ph/MeV LY under γ-ray excitation at room temperature. At higher Ce-concentration this material exhibit three decay time constants and they get longer with the increase of $Ce^{3+}$ content in the host. Overall, this study shows that TLLC is a promising scintillation material for radiation detection. Further study is underway for the improvement of this material.

Table I. Scintillation characteristics of TLLC: $Ce^{3+}$ crystals at room temperature.

| Ce-concentration (mole%) | Energy resolution at 662 keV ΔE/E (FWHM) % | Scintillation Light Yield (photons/MeV) | Decay Time |
|---|---|---|---|
| 0.5 | 7.8 | 23,500±2350 | 341 ns (83%), 995 ns (17%) |
| 1 | 5.6 | 27,000±2700 | 72 ns (8%), 366 ns (30%), 1.5 µs (62%) |
| 3 | 8.4 | 24,500±2450 | 78 ns (13%), 472 ns (32%), 2.6 µs (55%) |
| 5 | 7.0 | 25,000±2500 | 93 ns (13%), 595 ns (41%), 3.4 µs (46%) |

**Acknowledgment**

These investigations have been supported by the National Research Foundation of Korea (NRF) funded by the Ministry of Science and Technology, Korea (MEST) (No. 2015R1A2A1A13001843).

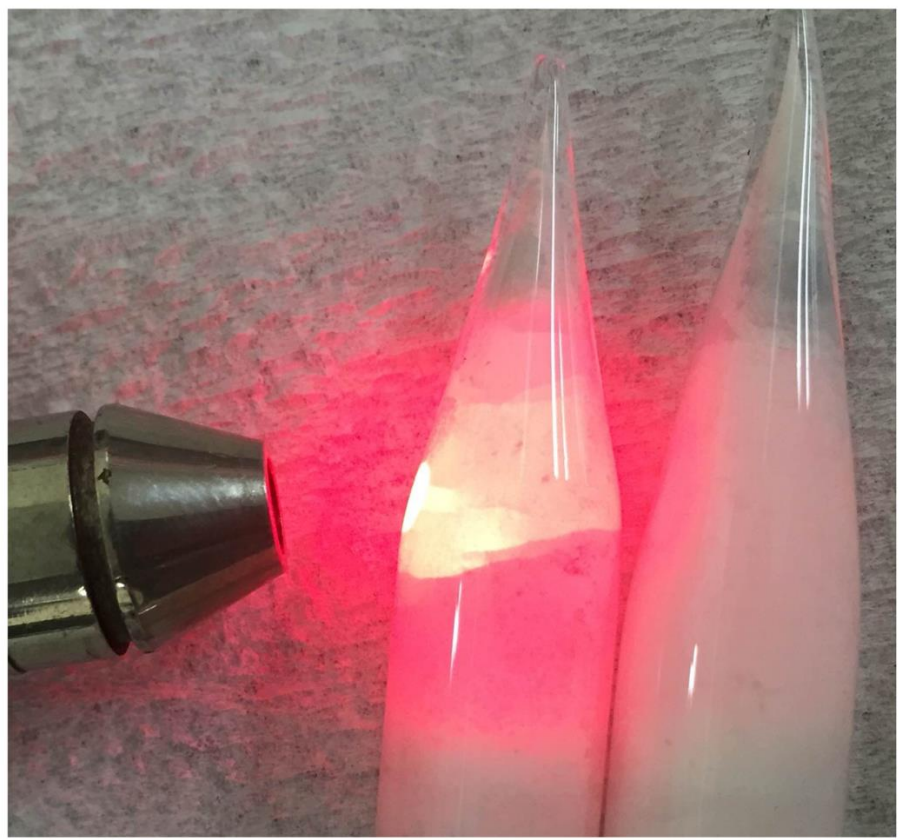

Figure 1. A photograph of the as grown samples of TLLC single crystals.

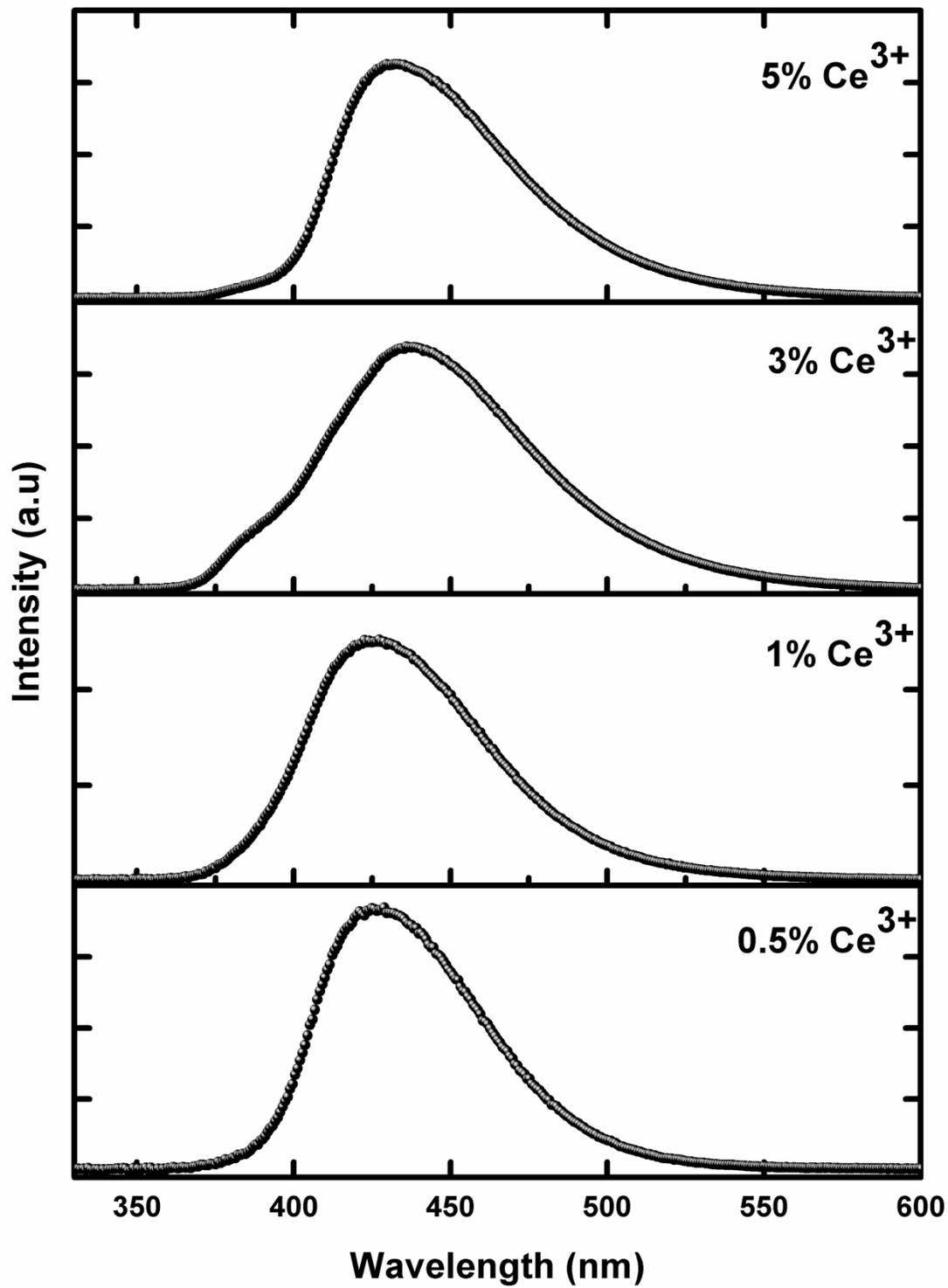

Figure 2. X-ray excited emission spectra at room temperature of TLLC: x $Ce^{3+}$ (x= 0.5%, 1%, 3%, and 5%).

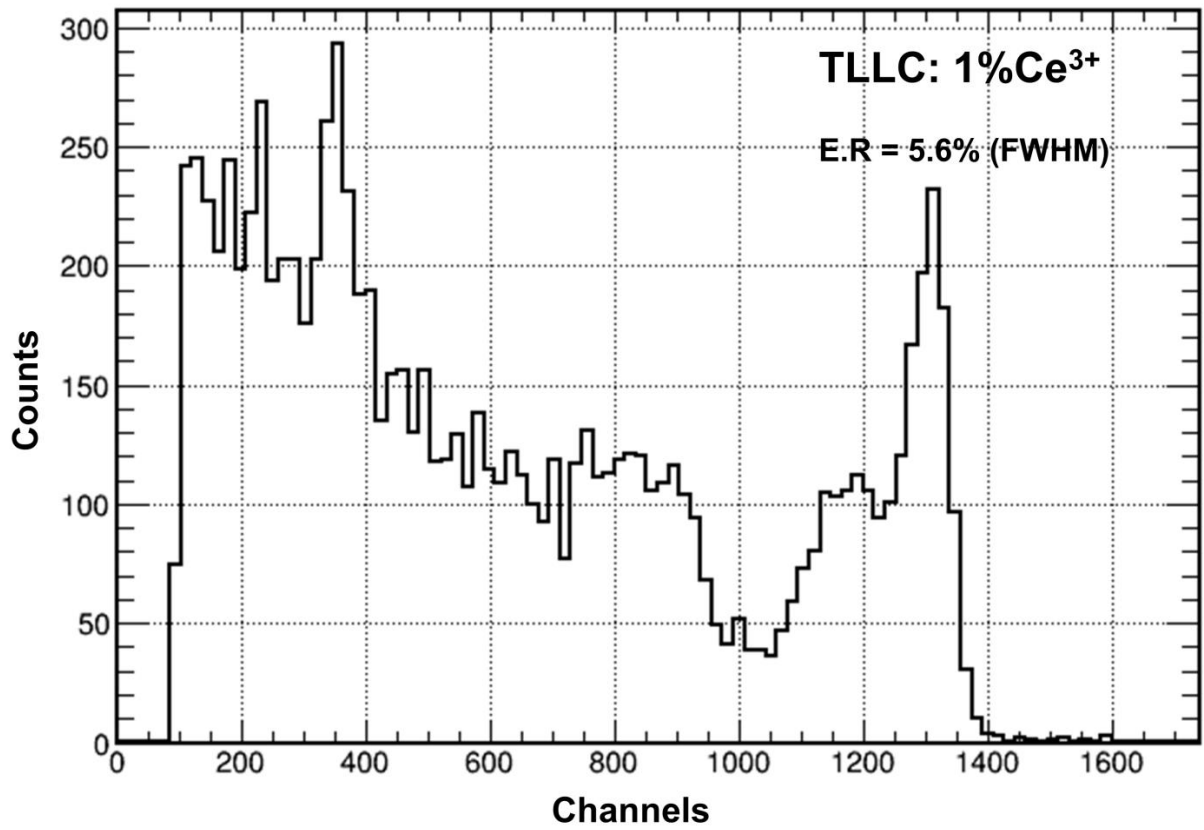

Figure 3. The scintillation pulse height spectrum of TLLC: 1% $Ce^{3+}$ irradiated with γ-rays from a $^{137}Cs$ source.

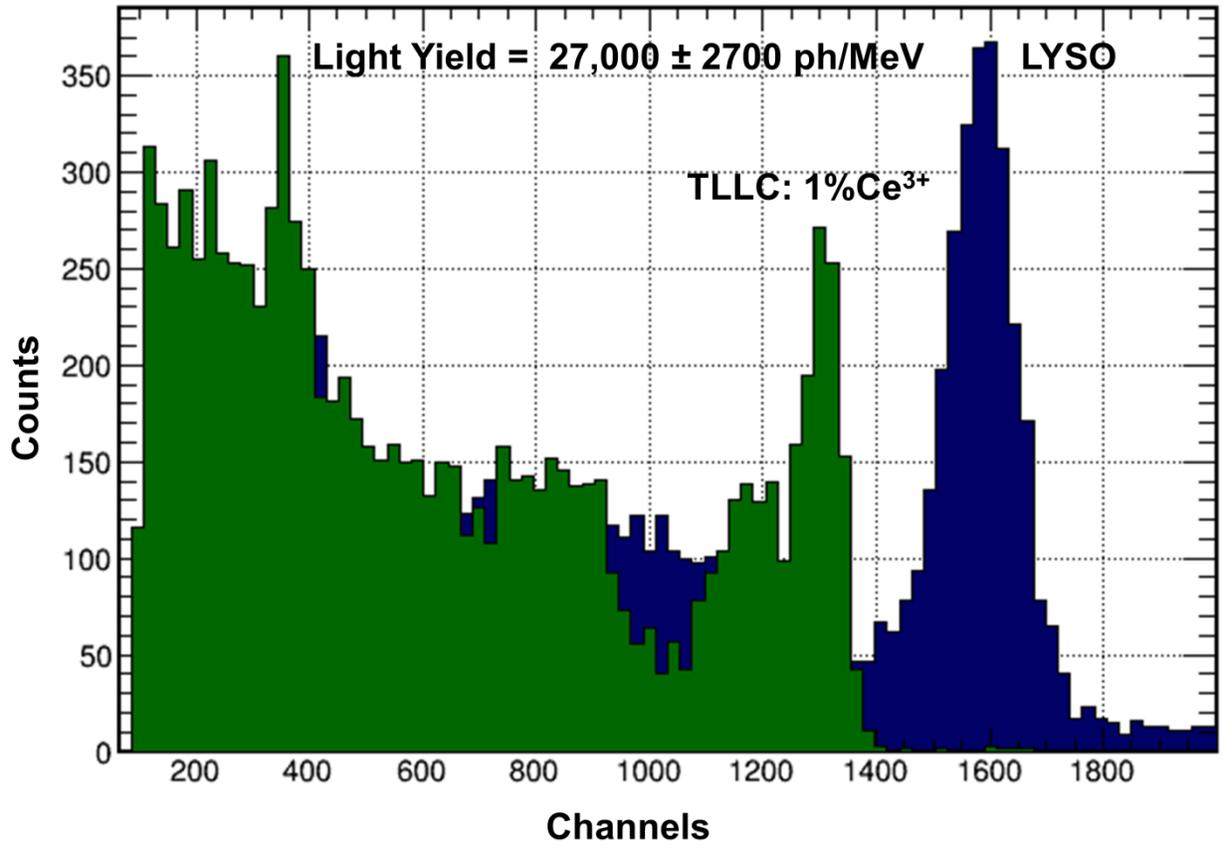

Figure 4. Pulse height spectra of TLLC: 1% $Ce^{3+}$ and LYSO: $Ce^{3+}$ crystals excited with γ-rays from a $^{137}$Cs source. The photopeak position is proportional to the light yield.

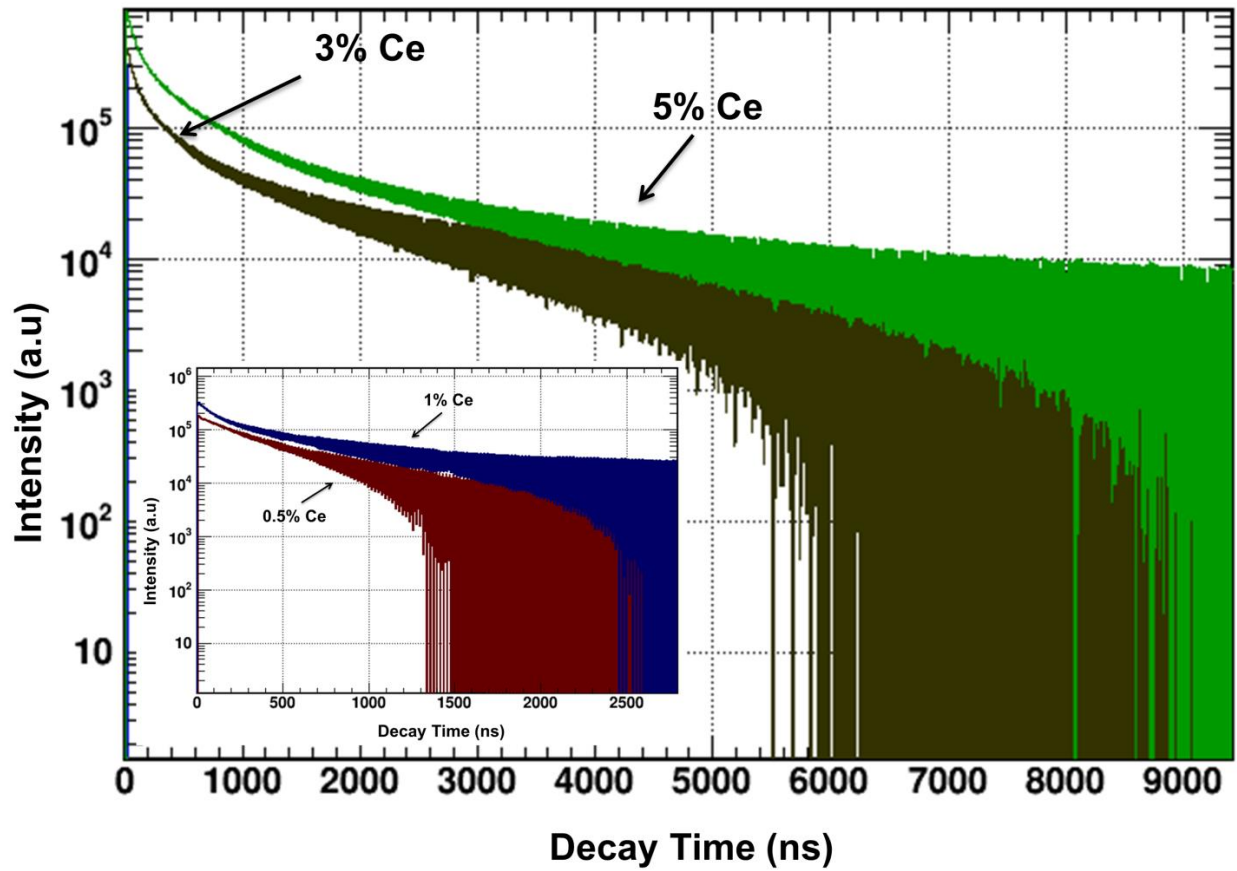

Figure 5. Time profiles of TLLC crystals with 0.5%, 1%, 3% and 5% Ce concentrations measured under $^{137}$Cs γ-rays excitation at room temperature.